\def\r{{\mathbf r}_i}
\def\k{{\mathbf k} }
\def\n{{\noindent}}

\def\S{{\mathbf S}}
\def\mbf{\mathbf }

\def\mns{\enskip -\enskip}

\documentclass[12pt,epsfig]{iopart}
\usepackage{graphicx}% Include figure files
\begin{document}
\jl{3}
\title [Lattice thermal conductivity of disordered NiPd and NiPt alloys]{Lattice thermal conductivity of disordered NiPd and NiPt alloys}
\author{Aftab Alam \footnote{E-mail : alam@bose.res.in} and 
Abhijit Mookerjee \footnote[1]{E-mail : abhijit@bose.res.in}}
\address{S.N. Bose National Centre for Basic Sciences,
 JD Block, Sector III, Salt Lake City, Kolkata 700098,
India}
\begin{abstract}
Numerical calculations of lattice thermal conductivity are reported for the binary alloys
 NiPd and NiPt. The present work is a continuation of an earlier paper by us 
\cite{lattice} which had developed a theoretical framework for the calculation of
 configuration-averaged lattice thermal conductivity and thermal diffusivity
 in disordered alloys. The formulation was based on the augmented space theorem
 \cite{ast} combined with a scattering diagram technique.
 In this paper we shall show dependence of the lattice thermal conductivity on a series of variables like phonon frequency, temperature and alloy composition. 
 The temperature dependence of $\kappa(T)$  and its relation to the measured
thermal conductivity is discussed.  The concentration dependence of $\kappa$ 
appears to justify the notion of a minimum thermal conductivity as discussed by 
Kittel, Slack and others \cite{kittel,slack}.  
 We also study the frequency and composition dependence of the thermal diffusivity 
averaged over modes. A numerical estimate of this quantity gives an idea about 
the location of mobility edge and the fraction of states in the frequency spectrum
which is delocalized.
\end{abstract}

\section{Introduction}
Lattice thermal conductivity of substitutionally disordered alloys  yield valuable information 
about the interactions of thermal excitations, phonons, with composition fluctuations
on their crystal lattice. Over the past few years numerous experimental studies 
\cite{expt1}-\cite{expt4} of the thermal conductivity of disordered alloys have
 provided considerable insight into the nature of their elementary excitations.

 The theory of lattice thermal conductivity for perfect crystals and ordered alloys has been set up on a rigorous basis. However the same is not true for disordered alloys. 
The presence of disorder results in scattering that not only depends on the 
impurity concentration but also crucially on both the relative masses and 
size difference between the constituent atoms. For large mass or size differences, 
the effect of disorder can be quite
unusual. Because of this, detailed comparison between theory and experiment on the basis of realistic models has not been very extensive. Model calculations are mostly based on 
mean-field approaches with diagonal disorder alone. Whereas in phonon problems 
essential off-diagonal disorder in the force constants cannot be dealt within  single site mean field approximations. Such disorder effects cannot be ignored in  realistic calculations.

 When we come to comparison with experiment, however, we face a different kind
of difficulty. 
In any experiment, the measured thermal conductivity $\kappa$ consists of the sum 
of an electronic component $\kappa_{e}$ and a lattice component $\kappa_L$
and  $\kappa=\kappa_{e}\ +\ \kappa_L$. Assuming the thermal analogue of Matthiessen's 
rule to be valid, the electronic thermal resistivity $W_{e}=1/\kappa_{e}$ is given by 
the sum of an ideal resistivity  and an impurity or residual term. It is often assumed 
that the ideal resistivity remains unaltered by alloying and can be obtained from 
the measurements on pure metals. The residual resistivity $W_r$ can be calculated with 
the help of Widemann-Franz law $W_{r}=\rho_{0}(T)/(L_{0} T)$, where $L_{0}$ is the 
Lorenz number and $\rho_{0}$(T) is determined by measuring the electrical resistivity 
of the alloy at several temperatures. The lattice component $\kappa_L$ can then be 
separated out from the observed conductivity $\kappa$. Overall what we would like 
to convey from these details  is that a direct measurement of the lattice component 
of the thermal conductivity is not feasible. There always exists certain assumptions 
behind the calculation of $\kappa_{e}$ and hence it is not possible always to obtain 
 reliable estimates for this quantity which consequently affects the separation of 
$\kappa_L$ from the observed conductivity $\kappa$.

In an earlier communication \cite{lattice}(hereafter referred to as AM)  we had developed a formalism for the calculation of configuration averaged lattice thermal conductivity and thermal diffusivity for a disordered binary alloy. Unlike single-site mean-field 
 approaches, this formulation has no restrictions on what kind of substitutional 
disorder can be studied. It explicitly takes into account fluctuations in masses, force constants and heat currents between different nuclei. It also maintains, on
the average, the sum rule between the diagonal and off-diagonal parts of the
dynamical matrix which eliminates the translational mode.
 We had shown that the dominant effect of disorder is to renormalize each propagator as well as the current terms in the Kubo formula. 

The purpose of this article is to implement the formulation (AM) for a detailed numerical study of the configuration averaged lattice conductivity and thermal diffusivity
 for disordered NiPt and NiPd alloys. The choice of the alloy systems is not arbitrary. We have chosen these alloys in order to propose a systematic study of the effects of mass disorder and strong force constant disorder. In the NiPd alloys, mass disorder is much larger than the force constant disorder. However both kinds of disorder dominate in the NiPt alloys. Moreover in both the two alloy systems there is a large size mismatch between the constituents. This indicates  that the standard single site mean field theories would be inadequate to capture these effects. These alloy systems are therefore ideal for the illustration of  the advantages of the augmented space block recursion method proposed by us \cite{am2}. Theoretically there have been many attempts \cite{theo1}-\cite{theo4} to develop
 an adequate approximation for understanding the lattice thermal conductivity of metals and alloys. The majority were based on model calculations either for perfect crystals or ordered alloys. 
One of the most successful mean-field approximation was the coherent potential approximation (CPA) \cite{CPA} in combination with the appropriate Kubo formula. The CPA is a single site mean-field theory capable of dealing only with mass disorder. There have been many attempts to generalize the CPA in order 
to treat both off-diagonal and environmental disorder. In many cases
these approximations did not satisfy the necessary herglotz analytic 
 and lattice translational invariance properties of the configuration averaged Green functions.
 Two very similar approaches, both based on the augmented space theorem,
have been recently proposed : the itinerant cluster CPA (ICPA) \cite{leath1}
and the augmented space recursion \cite{am1}. These have been used to 
study  the lattice dynamics of Ni$_{55}$Pd$_{45}$, Ni$_{50}$Pt$_{50}$ and Ni$_{88}$Cr$_{12}$ alloys. In a more generalized context of inelastic neutron scattering in disordered alloys, an augmented space block recursion (ASBR) has also been proposed recently \cite{am2}.
The ASBR  calculates the full Green and  Self-energy matrices instead of their diagonal entries alone. These are required for obtaining the response functions.

In this paper, we shall make use of the ASBR technique to implement the theoretical formulation for the lattice thermal conductivity developed earlier by us. We shall study the dependence of lattice thermal conductivity on phonon frequency as well as on temperature in a series of disordered NiPd and NiPt alloys covering the full range of alloy compositions. 

As far as the temperature dependence of lattice conductivity is concerned, our results follow a general trend. We shall also discuss how the lattice conductivity behave as a function of concentration at several fixed temperatures. The notion of a minimum thermal conductivity will be justified. It will be shown that low temperature resonant modes considerably decrease the conductivity. An idea about the location of the mobility edge will be discussed from the phonon-frequency dependence of the mode averaged harmonic diffusivity. This dependence also gives a rough idea about what fraction of the states across the frequency spectrum are
delocalized and therefore can carry current. The concentration dependence of the harmonic diffusivity will also be displayed.

\section{Theoretical formulation}

In the next two subsections, we shall apply the formalism introduced in our
earlier paper (AM) to study NiPd and NiPt alloys across the alloy
composition range. 
In (AM) we have discussed the theoretical formalism in great detail. We shall
present only the main pointshere and leave the reader to refer to (AM) for greater
detail.

The derivation of a Kubo formula for thermal conductivity requires an additional statistical hypothesis, which states that a system in steady state has a space dependent {\sl local} temperature $T({\r}) = [\kappa_B \beta({\r})]^{-1}$. 
 The matrix element of the heat current in the basis of the eigenfunctions of the Hamiltonian is given by~:

\[
{\mbf S}_{\gamma\gamma^\prime}^\mu (\k)\ =\ \frac{h}{2}\ \left(\rule{0mm}{3mm} \nu_{\k\gamma}+\nu_{\k\gamma^\prime}\right)\ {\mbf v}^\mu_{\gamma\gamma^\prime}(\k) ,
\]

\n where, the phonon group velocity ${\mbf v}_{\gamma\gamma^\prime}(\k)$ is given by

\[
{\mbf v}_{\gamma\gamma^\prime} 
  =  \frac{1}{2\sqrt{\nu_{\k\gamma}\nu_{\k\gamma^\prime}}}\                                 
\sum_\mu\sum_\nu \epsilon^\mu_\gamma(\k)\ {\mbf \nabla}_\k D^{\mu\nu}(\k) \ \epsilon^\nu_{\gamma^\prime}(\k)
\]

\n here $\gamma,\gamma^\prime$ label the various modes of vibration, $\nu_{\k\gamma},\nu_{\k\gamma^\prime}$ are their frequencies,  $\epsilon^\mu_\gamma(\k), \epsilon^\nu_{\gamma^\prime}(\k)$ are the polarization vectors and $D^{\mu\nu}(\k)$ is the Fourier transform of mass scaled dynamical matrix.

We shall consider the case where the temperature gradient is uniform within the system. 
The Kubo formula then relates the linear heat current response to the temperature gradient  field 

\[ \langle S^{\mu}(t) \rangle = - \sum_{\nu}\int_{-\infty}^{\infty} \ dt'\ \kappa^{\mu\nu}(t-t')\  
{\mbf \nabla}^\nu \delta T(t),  \]
\noindent where

\[ 
\kappa^{\mu\nu}(\tau) =  \Theta(\tau)\ \frac{1}{T}\int_0^\beta\  d\lambda \langle  S^{\mu}(-i\hbar\lambda),S^{\nu}(\tau) \rangle ,
\]

$\Theta(\tau)$ is the Heaviside step function,
and
\[
S(-i\hbar\lambda)\ =\ e^{\lambda H}\ S\ e^{-\lambda H}.
\]

\n $\langle\ \rangle$ on the right-hand side of the above equation denotes thermal averaging over 
states in the absence of the temperature gradient. The above equation can be rewritten in the form of a Kubo-Greenwood expression 

\begin{eqnarray}
\kappa^{\mu\nu}(\nu, T)  = \ \kappa^{\mu\nu}_I(\nu, T) \ +\ \kappa^{\mu\nu}_{II}(\nu, T) \nonumber\\
\phantom{x} \nonumber\\
\fl \kappa^{\mu\nu}_I (\nu, T) = \frac{\pi}{T}\ \int \frac{d^3\k}{8\pi^3}\ \sum_\gamma\sum_{\gamma^\prime\ne\gamma}\ \frac{\langle n_{\k\gamma^\prime}\rangle-\langle n_{\k\gamma}\rangle}{h(\nu_{\k\gamma}-\nu_{\k\gamma^\prime})} 
\ {\mbf S}^\mu_{\gamma\gamma^\prime}(\k){\mbf S}^\nu_{\gamma^\prime\gamma}(\k) \ \delta(\nu_{\k\gamma}-\nu_{\k\gamma^\prime}-\nu) \nonumber\\
\phantom{x}\\
 \fl \kappa^{\mu\nu}_{II}(\nu, T) = \frac{1}{\kappa_B T^2} \left[ \rule{0mm}{4mm}\left\{ \int\frac{d^3\k}{8\pi^3} \sum_\gamma \langle n_{\k\gamma}\rangle\ {\mbf S}^\mu_{\gamma\gamma}(\k)\right\} 
 \left\{ \int\frac{d^3\k}{8\pi^3}\sum_\gamma\langle n_{\k\gamma}\rangle\ {\mbf S}^\nu_{\gamma\gamma}(\k)\right\}\right.\nonumber\\
 - \kappa_B T \int\frac{d^3\k}{8\pi^3}\ 
 \left.\sum_\gamma \frac{\partial\langle n_{\k\gamma}\rangle }
{\partial(h\nu_{\k\gamma})}\ S^\mu_{\gamma\gamma}(\k)\ S^\nu_{\gamma\gamma}(\k) \right]\delta(\nu) 
\end{eqnarray}
where $\langle n_{\k\gamma}\rangle = (e^{\beta h\nu_{\k\gamma}}-1)^{-1}$ is the equilibrium Bose Einstein distribution function and T is the absolute temperature.

The first expression is for inter-band transitions, while the second expression is for intra-band transitions.
For an isotropic response, we can rewrite the first expression as 

\[
 \fl \kappa_I(\nu, T)= \frac{\pi}{3T}\sum_{\mu} \int d\nu^\prime \int\frac{d^3\k}{8\pi^3} \sum_{\gamma\gamma^{\prime}}
\widehat{\mbf S}^\mu_{\gamma\gamma^\prime}(\k, T)
\ \widehat{\mbf S}^\mu_{\gamma^{\prime} \gamma}(\k, T) \delta(\nu^\prime - \nu_{\k\gamma^\prime})\delta(\nu^\prime+\nu-\nu_{\k\gamma}) 
\]

\n where 

\[
\widehat{\mbf S}^\mu_{\gamma\gamma^\prime}(\k,T)\ =\ \sqrt{\left|\frac{\langle n_{\k\gamma^\prime}\rangle-\langle n_{\k\gamma}\rangle}{h(\nu_{\k\gamma}-\nu_{\k\gamma^\prime})}\right|}\ {\mbf S}^\mu_{\gamma\gamma^\prime}(\k).
\]

We may rewrite the above equation as 

\begin{eqnarray*}
\fl \kappa_I(\nu,T) = \frac{1}{3\pi T} \sum_{\mu} \int d\nu^\prime\int\frac{d^3\k}{8\pi^3} \mbox{Tr} \left[\rule{0mm}{4mm} \widehat{\mbf S}^{\mu}(\k,T)\Im m \{ {\mbf G}(\k,\nu^\prime)\}
 \widehat{\mbf S}^{\mu}(\k,T) \Im m\{ {\mbf G}(\k,\nu^\prime+\nu)\} \right]
\end{eqnarray*}

The operator {\bf G}($\nu$) is the phonon Green operator $(M\nu^2{\mbf I}-{\mbf\Phi})^{-1}$. The Trace is invariant in different representations. For crystalline systems, usually the Bloch basis 
$\{\vert {\mbf k},\gamma\rangle\}$ is used. For disordered systems, prior to configuration averaging,
  it is more convenient to use the basis $\{\vert \k,\alpha\rangle\}$, where $\k$ is the reciprocal vector and $\alpha$
represents the coordinate axes directions.  We can transform from the mode basis to the coordinate basis by
using the transformation matrices $\Upsilon_{\gamma\alpha}(\k) =\ \epsilon^\alpha_\gamma(\k)$. For example 

\[ \widehat{\mbf S}_{\alpha\beta}^\mu(\k,T) \ =\     \Upsilon^{-T}_{\alpha\gamma}(\k)\ \widehat{\mbf S}^\mu_{\gamma\gamma^\prime}(\k,T)\ \Upsilon^{-1}_{\gamma^{\prime}\beta}(\k).
\]

\noindent If we define

\begin{equation} 
{\mbf \kappa}(z_1,z_2)  = \int\frac{d^3\k}{8\pi^3}\ \mbox{Tr} \left[\ \rule{0mm}{4mm}\widehat{\mbf S}\  {\mbf G}(\k,z_1)\ \widehat{\mbf S}\ {\mbf G}(\k,z_2)\right] .
\label{eq3}
\end{equation}

\noindent  then the above equation becomes,
\begin{eqnarray}
 \fl\kappa(\nu,T) =  \frac{1}{12\pi T}\ \sum_{\mu}  \int d\nu^\prime \ \left[\rule{0mm}{4mm} {\mbf \kappa}^{\mu\mu}(\nu^{\prime -},\nu^{\prime +}+\nu)
 + {\mbf \kappa}^{\mu\mu}(\nu^{\prime +},\nu^{\prime -}+\nu) \right.\nonumber\\
  \left. - {\mbf \kappa}^{\mu\mu}(\nu^{\prime +},\nu^{\prime  +}+\nu) - {\mbf \kappa}^{\mu\mu}(\nu^{\prime -},\nu^{\prime  -}+\nu)\right]
\label{ref2}
\end{eqnarray}

\noindent where

\[ 
f(\nu^+) = \lim_{\delta\rightarrow 0} f(\nu+i\delta),\phantom{xx}
f(\nu^-) = \lim_{\delta\rightarrow 0} f(\nu-i\delta).
\]

We have used the herglotz analytic property \cite{am1} of the Green operator   

\[
{\mbf G}(\nu+i\delta) = \Re e\left[\rule{0mm}{3mm}{\mbf G}(\nu)\right] \mns  i\ \mbox{sgn}(\delta)\ \Im m\left[\rule{0mm}{3mm}{\mbf G}(\nu)\right]. 
\]

For disordered materials, we shall be interested in obtaining the configuration averaged  response functions. This will require
the configuration averaging of quantities like $\kappa(z_1,z_2)$. This averaging
procedure we have discussed in detail in (AM) :

\begin{equation}
\ll \kappa(z_1,z_2)\gg \ =\ \ll \kappa_{(1)}(z_1,z_2)\gg + \ll \Delta\kappa(z_1,z_2)^{\mathrm{ladder}}\gg
\label{eq5}
\end{equation}

 The first term in the right hand side stands for the disorder induced corrections where the disorder scattering renormalizes the phonon propagators as well as the heat currents. In this term, corrections to the heat current is related to the self-energy of the propagators. The second term of Eq. (\ref{eq5}) includes the vertex corrections due to the correlated propagation.

For a harmonic solid, a temperature independent mode diffusivity $D_{\gamma}$ is defined as 
\[
D^{\mu\nu}_{\gamma}(\k)=\pi \sum_{\gamma^{\prime}\ne\gamma}\frac{1}{\nu_{\k\gamma}^{2}}{\mbf S}^{\mu}_{\gamma\gamma^{\prime}}(\k){\mbf S}^{\nu}_{\gamma^{\prime}\gamma}(\k)\ \delta(\nu_{\k\gamma}-\nu_{\k\gamma^{\prime}}).
\]
This is an intrinsic property of the $\gamma$-th normal mode and provides an unambiguous criterion for localization.

The averaged thermal diffusivity (averaged over modes) is then given by
\[
{\bf D}^{\mu\nu}(\nu)=\frac{\displaystyle\int\frac{d^{3}\k}{8 \pi^{3}}\sum_{\gamma}D^{\mu\nu}_{\gamma} (\k)\delta(\nu - \nu_{\k\gamma})}{\displaystyle\int\frac{ d^{3}\k}{8 \pi^{3}}\sum_{\gamma}\delta(\nu - \nu_{\k\gamma})}
= \frac{D^{\mu\nu}_{tot}(\nu)}{{\displaystyle\int\frac{ d^{3}\k}{8 \pi^{3}}\sum_{\gamma}\delta(\nu - \nu_{\k\gamma})}}
\]

 Assuming isotropy of the response, we can rewrite the numerator of above equation as

\begin{eqnarray*}
\fl D^{\mu\mu}_{tot}(\nu)=\pi\int d\nu^{\prime}\int \frac{d^{3}\k}{8 \pi^{3}}\sum_{\gamma}\sum_{\gamma^{\prime}}\widehat{S}^{\mu}_{\gamma\gamma^{\prime}}(\k)\widehat{S}^{\mu}_{\gamma^{\prime}\gamma}(\k)
 \delta(\nu^{\prime}-\nu_{{\k}\gamma^{\prime}})\delta(\nu_{{\k}\gamma}-\nu^{\prime})\delta(\nu-\nu_{{\k}\gamma})
\end{eqnarray*}
\noindent where
\[
\widehat{S}^{\mu}_{\gamma\gamma^{\prime}}(\k)\ =\ \frac{1}{\nu_{\k \gamma}}{\mbf S}^{\mu}_{\gamma\gamma^{\prime}}(\k).
\]
We may again rewrite the above equation for $D_{tot}$ as
\begin{eqnarray*}
\fl D^{\mu\mu}_{tot}(\nu)=\frac{1}{\pi^{2}}\int d\nu^{\prime}\int \frac{d^{3}\k}{8 \pi^{3}} \mbox{Tr} \left[ \Im m \{{\mbf G}(\k,\nu^{\prime})\} \widehat{\mbf S}^{\mu}(\k) 
  \Im m \{{\mbf G}(\k,\nu^{\prime})\}\widehat{\mbf S}^{\mu}(\k)\Im m \{{\mbf G}(\k,\nu)\}  \right]
\end{eqnarray*}

 The averaged thermal diffusivity can then be expressed as (for an isotropic response)

\begin{eqnarray}
{\mbf D}(\nu)=\frac{1}{3}\sum_{\mu}D^{\mu\mu}(\nu)
=\frac{\pi}{3}\frac{\sum_{\mu}{\mbf D}_{tot}^{\mu\mu}(\nu)}{\displaystyle\int \frac{d^{3}\k}{8\pi^{3}}\mbox{Tr}\left[\Im m \{{\mbf G}(\k,\nu)\}\right]}
\end{eqnarray}
For disordered material, we shall be interested as before in obtaining the configuration averaged thermal diffusivity. The configuration averaged thermal diffusivity can be expressed (to a 1st order approximation) in the form

\begin{equation}
\ll{\mbf D}(\nu)\gg =\frac{\pi}{3}\frac{\sum_{\mu}\ll{\mbf D}_{tot}^{\mu\mu}(\nu)\gg}{\displaystyle\int \frac{d^{3}\k}{8\pi^{3}}\mbox{Tr}\left[\Im m \ll{\mbf G}(\k,\nu)\gg\right]},
\end{equation}
\noindent where
\begin{eqnarray*}
\fl \ll D^{\mu\mu}_{tot}(\nu)\gg\ \simeq\ \frac{1}{\pi^{2}}\int d\nu^{\prime}\int \frac{d^{3}\k}{8 \pi^{3}} \mbox{Tr} \left[\rule{0cm}{4mm} \Im m \ll{\mbf G}(\k,\nu^{\prime})\gg \right.\\
\left. 
\ll \widehat{\mbf S}^{\mu}(\k)\ \Im m\ \{{\mbf G}(\k,\nu^{\prime})\} \ \widehat{\mbf S}^{\mu}(\k)\ \rule{0cm}{4mm} \Im m \{{\mbf G}(\k,\nu)\}\gg  \right]
\end{eqnarray*}

\section {RESULTS AND DISCUSSION}
 The details of numerical calculation for the two alloys of our interest in the present work are as follows : 
\begin{itemize}
\item We have carried out calculations on 501 $\nu$-points.
\item A small imaginary part of the frequency $\delta=$0.001 has been used for evaluating the Green matrix and Self-energy matrix in the augmented space block recursion \cite{am2}.
\item The calculation of lattice conductivity has been done at 40 temperatures.
\item For the Brillouin zone integration, 145 ${\bf k}$-points in the irreducible $1/48$-th of the zone produced well converged results.
\end{itemize}

\subsection{NiPd alloy : Strong mass and weak force constant disorder.}

  For a list of general properties of
 fcc Ni and Pd, we refer the reader to Ref. \cite{am1}. This particular alloy has already
 been studied experimentally by Farrell and Greig \cite{expt1} using conventional
 potentiometric techniques. But unfortunately their investigation was limited only to 
very dilute alloys in the temperature range 2-100 K. In an earlier communication 
 (AM) 
 we have already shown a comparison of our results for the temperature dependence of thermal conductivity with theirs for a dilute Ni$_{99}$Pd$_{01}$ alloy.

\begin{figure}[t]
\vskip 1cm
\centering
\includegraphics[width=9cm,height=6.7cm]{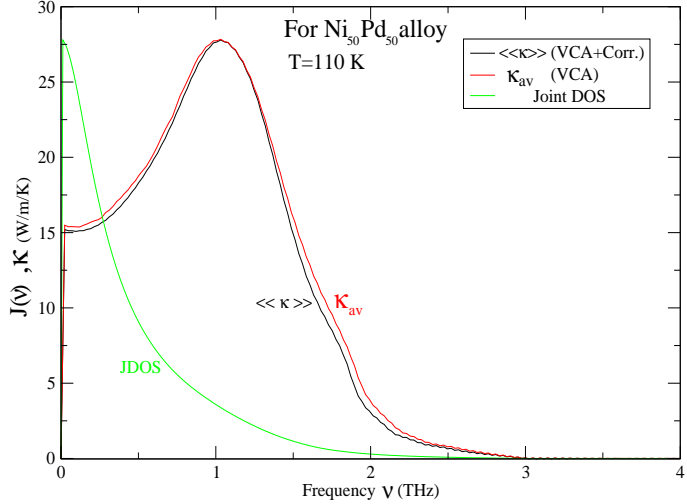}
\caption{(Color Online)\ Configuration averaged lattice thermal conductivity vs phonon frequency $\nu$ (THz) for  Ni$_{50}$Pd$_{50}$ disordered alloy. The red line and black line shows the conductivity using the average VCA current and effective current (consisting of average VCA current + disorder corrections + vertex correction) respectively. The green line in indicates the configuration averaged joint density of states. }
\label{com1}
\end{figure}

Our initial focus in the earlier communication \cite{lattice} was to calculate the effect of disorder scattering over an averaged medium. We had made use of the  
multiple scattering diagram technique and went beyond the framework of CPA
 which treats only diagonal disorder.  Rather we had applied the technique 
in a more generalized context with the inclusion of diagonal as well as
  off-diagonal disorder arising out of the disorder in the dynamical matrix.
 In an attempt to compare the effect of averaged heat current and 
disorder induced renormalized heat current on the lattice thermal conductivity, we have plotted Fig. \ref{com1}. 

Fig \ref{com1} shows the results for disordered $Ni_{50}Pd_{50}$ alloy. 
The black curve represents the lattice conductivity including all kinds of disorder induced corrections : e.g. corrections to the heat current and the vertex corrections, while the red curve stands for the same quantity but using averaged heat currects and without vertex corrections.
 The green curve in this figure  shows the scaled joint density of states. From the figure it is clear that the transition rate `${\mathbf \tau}$' is strongly dependent both on the initial and the final energies throughout the phonon frequency ($\nu$). That is
\begin{equation}
\kappa(\nu,T)\ne\vert{\mathbf{\tau}(\nu,T)}\vert J(\nu)\nonumber,
\end{equation}
where $J(\nu)$ is the joint density of states given by
\begin{eqnarray}
\fl J(\nu)\ =\ \int d\nu^\prime\int\frac{d^3\k}{8\pi^3}\ \mbox{Tr} \left[\rule{0mm}{4mm} \Im m \ll{\bf G}({\bf k},\nu^\prime)\gg
\Im m \ll{\bf G}({\bf k},\nu^\prime + \nu)\gg\right]
\label{JDOS}
\end{eqnarray}

 Figure \ref{com1}  shows that, the effect of disorder corrections to the current terms on the overall shape of lattice conductivity is rather small  in the  $Ni_{50}Pd_{50}$ alloy. 
We should note that the effect of disorder corrections to current and
the vertex corrections in these alloys become
negligible beyond  phonon frequencies $\simeq$ 2.5 THz.
For the  higher frequency modes the effect of scattering phenomenon 
is well described by the mean filed approximation. 
It is in  the low frequency region that configuration fluctuation effects beyond
the  mean-field becomes significant.

There is a very important feature in Fig. \ref{com1}  that still needs discussion : which is how to explain the origin of a dip in $\kappa(\nu)$ at the lowest energy $\nu =0$ ? A similar kind of dip has also been reported by Feldman {\it et.al.} \cite{feldmanetal} while studying amorphous Si and Si$_{1-x}$Ge$_{x}$ alloys. Their  $\kappa(\nu)$ have a small 
Lorenzian shaped dip centred at $\nu$=0. This  reflects the missing intraband conductivity  $\kappa^{II}$.
 This dip in $\kappa(\nu)$ stands at a small but finite value [$\nu\simeq$0]. The finiteness of the dip in $\kappa(\nu)$ is because of the fact that their calculation was based
 on a Kubo-Greenwood expression for the thermal conductivity with the delta functions in the expression broadened into a Lorenzian of small (but finite) width $\eta$. However in our case it is evident from the Fig. \ref{com1} that this dip in $\kappa(\nu)$ stands at $\kappa(\nu)\rightarrow 0$ at $\nu =0$. This is due to the simple reason that in our calculation the 
Lorenzian broadening has not been put in by hand, but it arises automatically 
 from the disorder effect on the crystalline spectral function $\Im m [{\mathbf G}({\bf k},\nu)]$ . Another reason for this difference in the position of dip in $\kappa(\nu)$ may be due to the fact that Feldman {\sl et.al.} carried out their calculation at a fixed wavevector
 ${\bf k}$, while we have  summed over the entire Brillouin zone.

The origin of this dip can also be explained by looking at the joint density of states (JDOS) represented by green line in Fig. \ref{com1}. This quantity has a dip near $\nu=0$ reminiscent of the dip in the $\kappa{(\nu)}$ curves. This indicates that a smooth
 convolution of two Green matrices ${\mathbf G}({\bf k},\nu^\prime)$
 and ${\mathbf G}({\bf k},\nu^\prime +\nu)$ (or two smooth densities 
of states obtained after summing ${\bf k}$ over the Brillouin zone),
as appeared in the expression (5) of \cite{lattice},  is  mainly responsible for such 
a sharp dip in the lattice conductivity at $\nu=0$. As discussed by Feldman${\it et.al}$,
 this dip at $\nu=0$ disappears as the system size $N\rightarrow\infty$. 
They have also suggested an appropriate method to eliminate this dip in a sensible 
manner, which allows us to extrapolate the $\kappa(\nu)$ curve from a value at $\nu > 0$ 
($\nu\simeq$0) to a value at $\nu$=0. This extrapolated value of $\kappa(\nu)$  at $\nu=0$ is nothing but the d.c. value of the lattice thermal conductivity $\kappa_{0}$. 
In an attempt to calculate ($\kappa_{0}$), we have obtained a value of  15.25 W/m/K for $Ni_{50}Pd_{50}$ alloy at T=110 K. 

\begin{figure}[h]
\vskip 1cm
\centering
\includegraphics[width=3.3in,height=4.5in]{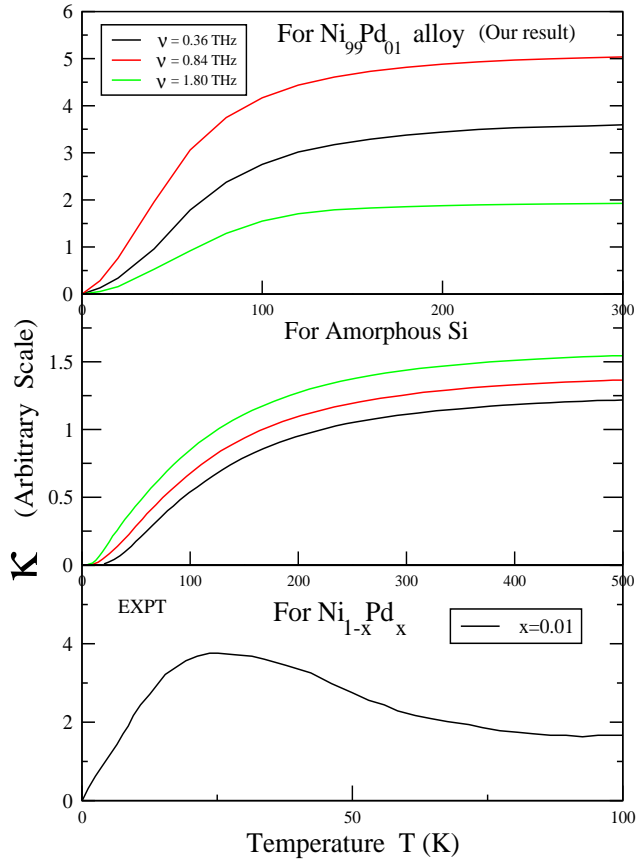}
\caption{(Color Online)\ Thermal conductivity vs temperature T(K) for NiPd alloys and Amorphous Si. The top panel shows our results on the lattice conductivity for Ni$_{99}$Pd$_{01}$ alloy at three different frequency cut-off $\nu$. The middle panel shows the lattice conductivity for amorphous Si \cite{feldmanetal} at three different cut-off frequency, while the panel at the bottom shows the experimental data \cite{expt1} for the total thermal conductivity (=\ lattice + electronic contribution) of the same Ni$_{99}$Pd$_{01}$ alloy. }
\label{fig1}
\end{figure}

\begin{figure}
\centering
\rotatebox{270}{\includegraphics[width=2.0in,height=4.0in]{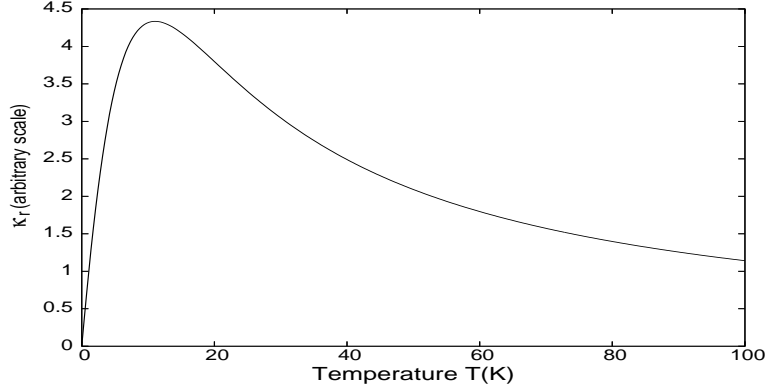}}
\caption{Residual or impurity contribution of the electronic part of the thermal
conductivity}
\label{fig2}
\end{figure}

 Direct comparison with the experimental data on these systems is difficult, because the experimental thermal conductivity also has a
component arising out of the contribution from electrons.
 Figure \ref{fig1} shows  the temperature dependence of lattice conductivity. The top panel shows our theoretical result for the $Ni_{99}Pd_{01}$ alloy at three different frequencies.
The bottom panel shows the experimental data \cite{expt1} on the total ( electronic and  lattice  ) thermal conductivity of the same 99-01 NiPd alloy.
Since the frequency is not mentioned in the
experimental data, we assume that it must be for low frequencies. The best comparison then  will be between the
middle (black) curve on the top panel and that in the bottom one. The two agree qualitatively, except at low temperatures where
we expect the electronic contribution to dominate. In order to understand
whether the deviation {\sl does arise} from the electronic contribution, we have compared the top panel
with the thermal
conductivity of amorphous-Si  \cite{feldmanetal}, shown in the middle panel. In a-Si the electrons near the Fermi level are
localized and hence cannot  carry any current. The contribution to thermal conductivity arises
from scattering due to configuration fluctuations in the amorphous material. Qualitatively we expect
the results to be similar to configuration fluctuation scattering in random alloys.
 Almost the entire contribution should come from
the phonons. The behaviour of the two panels are quite similar. The origin of the hump in the
experimental lattice conductivity can also be understood if we assume Widemann-Franz law and write
the residual part of the electronic contribution of the thermal resistivity as $\kappa_r =
L_0 T/\rho_0(T)$. Here $L_0$ is the Lorenz number and $\rho_0(T)$ is the electrical resistivity.
Assuming that the electrical resistivity behaves as $\rho_0(T)= A+BT+CT^2$ at low temperatures and
with a suitable choice of the parameters, this contribution does show a hump followed by a decreasing
behaviour flattening out at larger temperatures (see Fig \ref{fig2}). The sum of the contribution
shown in the top panel of Fig \ref{fig1} and that in Fig \ref{fig2} would lead to the experimental
behaviour shown in the bottom panel of Fig \ref{fig1}. This is a plausibility argument and needs to
be confirmed by a detailed calculation of the electronic contribution to the thermal conductivity.
\vskip 0.2cm

\begin{figure}[t]
\vskip 1cm
\centering
\includegraphics[width=10.0cm,height=6.0cm]{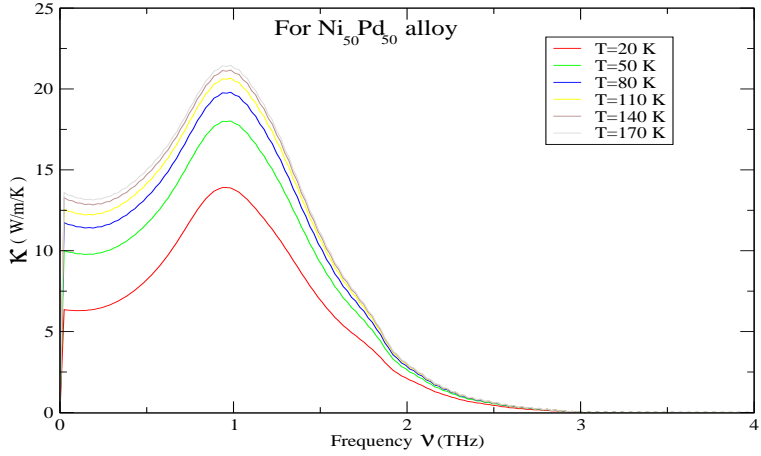}
\caption{(Color Online)\ The configuration averaged lattice thermal conductivity vs phonon frequency $\nu$ (THz) at different temperatures T for Ni$_{50}$Pd$_{50}$ alloy.}
\label{fig3}
\end{figure}

\begin{figure}[h]
\vskip 0.7cm
\centering
\includegraphics[width=10.0cm,height=6.0cm]{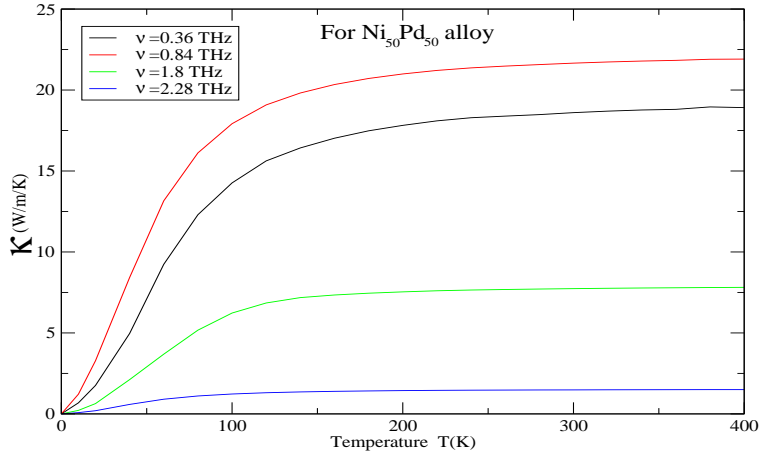}
\caption{(Color Online)\ The averaged lattice thermal conductivity vs temperature T(K) at various cut-off frequencies $\nu_{cutoff}$ for Ni$_{50}$Pd$_{50}$ alloy.}
\label{fig4}
\end{figure}
We shall now present the dependence of lattice thermal conductivity and thermal diffusivity on parameters such as phonon frequency ($\nu$), concentration ($x$) etc. In Fig. \ref{fig3}, we display the frequency dependence of lattice conductivity for Ni$_{50}$Pd$_{50}$  alloy at various temperatures evaluated using Eq. \ref{eq5}. The figure clearly shows the saturation  of lattice conductivity  as we proceed towards the higher temperatures. 
The d.c. value of the conductivity ($\kappa_0$), which is just  extrapolation of $\kappa(\nu)$ curve
 from a value at $\nu >0$ to a value at $\nu=0$, increases as we increase the temperature.  

The temperature dependence of lattice thermal conductivity for the Ni$_{50}$Pd$_{50}$ alloy at 
various phonon frequencies $\nu$ are shown in Fig. \ref{fig4}. It is qualitatively similar 
to  Ni$_{50}$Pt$_{50}$ alloy. The conductivity initially increases with temperature reflecting 
a quadratic T-dependence (in the low T-regime) and ultimately reaches a T-independent saturated
 value (at  higher temperatures).

\begin{figure}[t]
\vskip 1cm
\centering
\includegraphics[width=10cm,height=6.0cm]{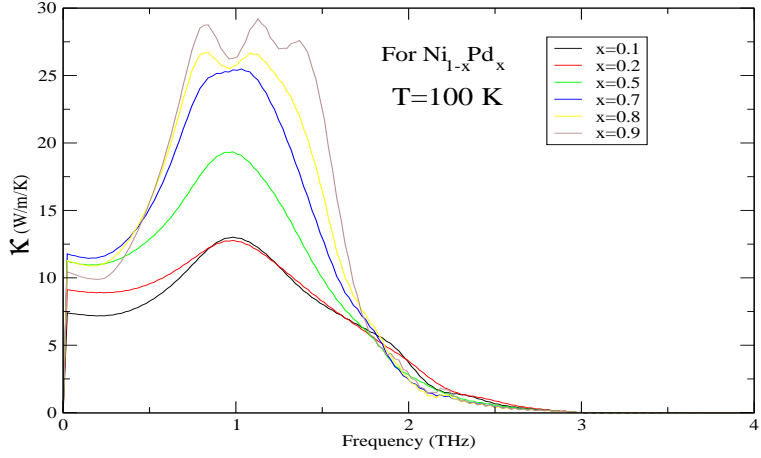}
\caption{(Color Online)\ Frequency dependence of lattice thermal conductivity for various alloys  Ni$_{1-x}$Pd$_{x}$ at T=100 K.}
\label{fig5}
\end{figure}
\begin{figure}[t]
\vskip 0.6cm
\centering
\includegraphics[width=9cm,height=5.5cm]{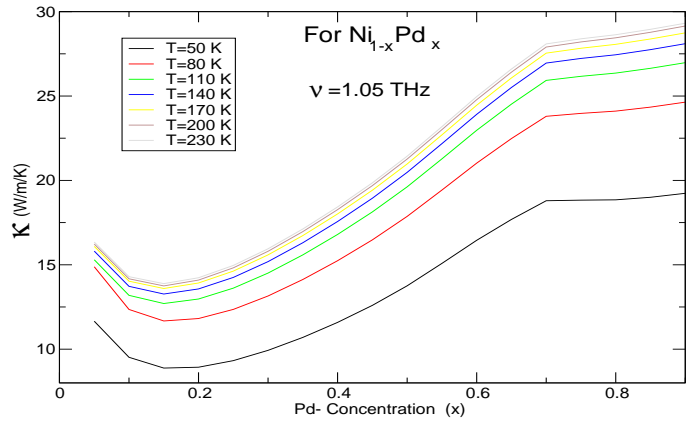}
\caption{(Color Online)\ Lattice thermal conductivity vs Pd-concentration for various temperature T at phonon frequency $\nu=1.05 THz$.}
\label{fig6}
\end{figure}

Figure \ref{fig5} shows the lattice conductivity  as a function of frequency
 at T=100 K for various alloy compositions. Comparing the results of Fig. \ref{fig5} 
with those of Fig. \ref{fig12} for NiPt, it is clear at a glance that the overall shape of frequency
 dependence of $\kappa$ for various alloys of Ni$_{1-x}$Pd$_{x}$ looks similar, however
for $x$=0.8 and $x$=0.9 extra structure appears in the frequency dependence.
 Similar behaviour has also been observed for $x$=0.9 in Ni$_{1-x}$Pt$_{x}$ alloy in Fig. \ref{fig12}.
We believe that this behaviour for Ni$_{1-x}$Pd$_{x}$ alloy at $x$=0.8  and 0.9 
 may be due to the strong disorder in masses, the effect of which becomes important in the 
two dilute limit alloys. Such an anomalous behaviour is also reflected in the concentration 
dependence of lattice conductivity. This is shown in Fig. \ref{fig6}, which plots 
the lattice conductivity vs Pd-concentration ($x$) at a fixed frequency $\nu$=1.05 THz for 
various temperatures. 

\begin{figure}[h]
\centering
\includegraphics[width=13cm,height=14cm]{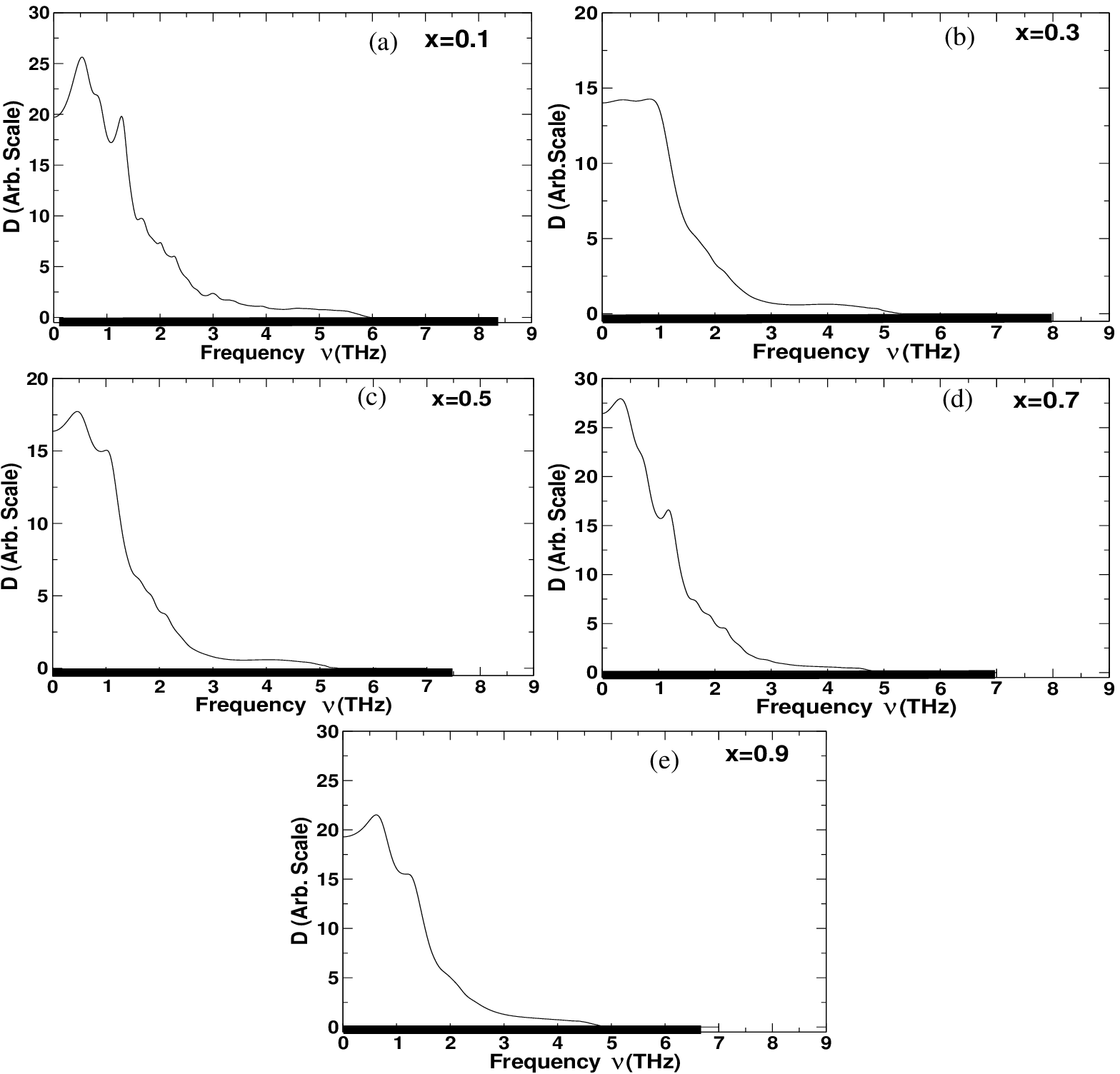}
\caption{The configuration averaged thermal diffusivities $D(\nu)$ for Ni$_{1-x}$Pd$_{x}$ alloys. (a)\ x=0.1;\ (b)\ x=0.3;\ (c)\ x=0.5;\ (d)\ x=0.7;\ (e)\ x=0.9.The broad line on the frequency axis shows the extent of the vibrational spectrum.}
\label{fig7}
\end{figure}

\begin{figure}
\centering
\includegraphics[width=9.0cm,height=10.0cm]{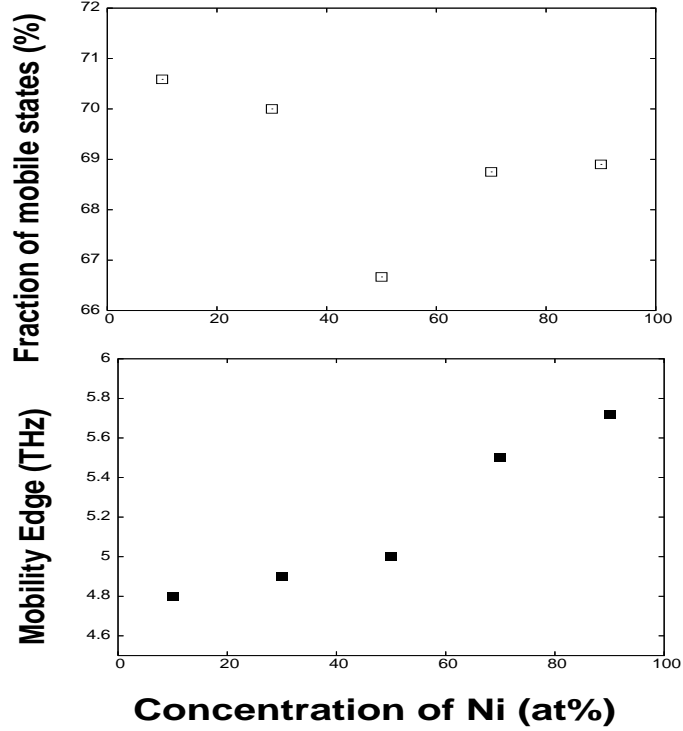}
\caption{The position of the mobility edge (bottom) and the percentage of
mobile phonon states (top) as a function of the alloy composition for NiPd alloy.}
\label{fig8}
\end{figure}

The thermal diffusivities D($\nu$) are important because the effect of disorder is often manifested in them more directly than in the conductivities. Not only that thermal diffusivity also gives an approximate idea about the location of mobility edge as well as the fraction of delocalized states. In Fig. \ref{fig7} we display the thermal diffusivity $D(\nu)$ vs frequency for various compositions. The extent of the phonon frequency spectrum is shown by the broad lines. The density of states is non-zero across this spectrum range.  The first thing to note is that, the region of large diffusivity in the five sets of  alloys at the higher frequency side is not the same. In other words the weakly defined hump in the lower as well as higher frequency side are located at different positions for different alloys. The low frequency maximum in diffusivity is a minimum around the 50-50 composition where the disorder scattering is the maximum. Above 2.8 THz, there is a smooth decrease of diffusivity approximately linear in frequency \ D($\nu)\propto (\nu_{c}-\nu)^{\alpha}$\ , with the critical exponent $\alpha\simeq 1$\,\ and a critical frequency $\nu_{c}$ where $D(\nu)$ vanishes to within a very small level of noise. The allowed phonon states beyond this frequency must be due to localized phonon modes. The critical frequency $\nu_{c}$ locates the mobility edge above which the diffusivity is strictly zero in the infinite size limit. Once the mobility edge is located, the fraction of de-localized states may be estimated directly.  It is clear from the figure that location of mobility edge varies with composition. Consequently the percentage of de-localized states available for thermal conduction in the system also varies with composition. An inspection of Fig. \ref{fig7} determines the location  of the mobility edges ($\nu_{c}$) for the five different compositions. The following figure \ref{fig8} which shows the position of the mobility edge and the percentage of mobile phonon states in the spectrum as a function of the composition is quite illustrative.

\begin{figure}[t]
\centering
\vskip 0.7cm
\includegraphics[width=10cm,height=6.0cm]{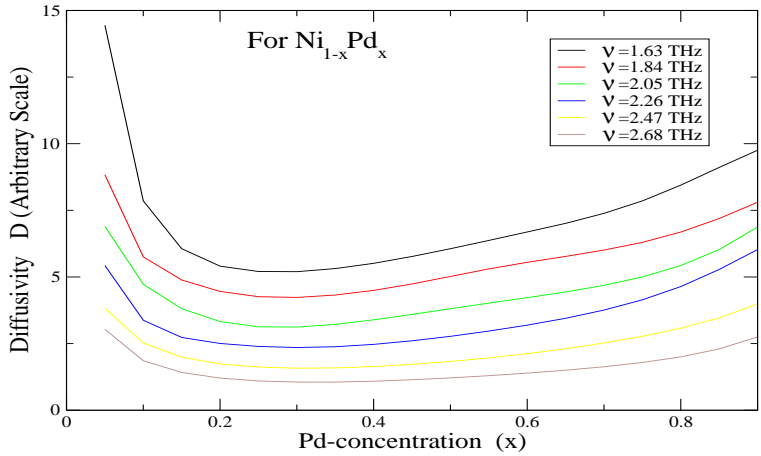}
\caption{(Color Online)\ The averaged thermal diffusivity $D(\nu)$ vs Pd-concentration at various cut-off frequencies $\nu_{cutoff}$ for Ni$_{1-x}$Pd$_{x}$ alloy. }
\label{fig9}
\end{figure}
The maximum percentage of localized states occur at 50-50 composition where
we expect disorder scattering to be a maximum. The mobility edge moves to
higher frequencies as the concentration of Ni increases, but so does the
band width of the phonon spectrum. The minimum percentage of mobile phonon
states available for thermal conduction occurs, as expected, at around the 50-50 composition.
A similar behaviour has also been discussed by Feldman {\it et.al} \cite{feldmanetal} while 
studying the effects of mass disorder on various Si$_{1-x}$Ge$_{x}$ alloys. 

Figure \ref{fig9} shows the concentration dependence of thermal diffusivity for Ni$_{1-x}$Pd$_{x}$ alloy at various frequencies ($\nu$). As expected from our earlier discussion, the minumum diffusivity
occurs around the 50-50 composition, where disorder scattering is maximum. The curves have asymmetry around $x=0.5$ which decreases with increasing frequency. This asymmetry reflects a similar asymmetry in the thermal conductivity as a function of composition.

\subsection{ NiPt alloy : Strong mass and force constant disorder.}
We refer the reader to a previous article \cite{am1} by us for some of the basic properties  of fcc Ni and Pt which is relevant for our present calculation. It has been our experience \cite{am1,am2} that the effect of disorder in NiPt alloy is more dramatic than NiPd. For instance the appearance of sharp discontinuities observed in the dispersion where we have resonance states and consequent increase in the line-width \cite{am1}. 

\begin{figure}
\vskip 0.8cm
\centering
\includegraphics[width=9cm,height=6.5cm]{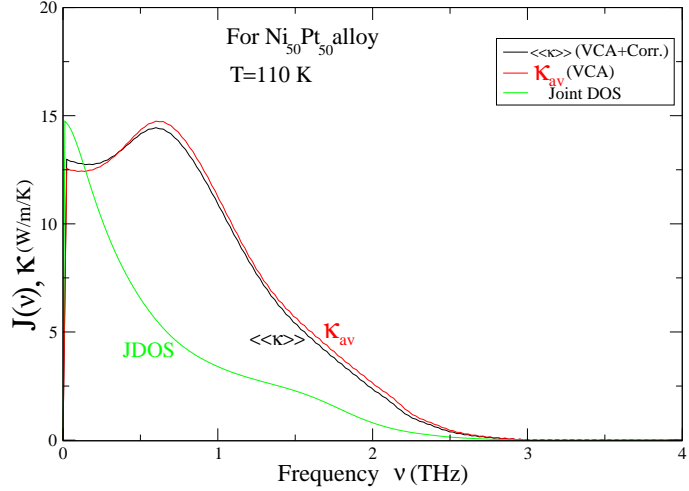}
\caption{(Color Oncurve)\ Configuration averaged lattice thermal conductivity vs phonon
 frequency $\nu$ (THz) for Ni$_{50}$Pt$_{50}$ disordered alloy. The red curve and black curve shows the conductivity using the average VCA current and effective current (consisting of average VCA current + disorder corrections+ vertex correction) respectively. The green curve  indicates the configuration averaged joint density of states. }
\label{com2}
\end{figure}

Figure \ref{com2} shows the results for disordered $Ni_{50}Pt_{50}$ alloy. 
As before, the black curve represents the lattice conductivity including 
all kinds of disorder induced corrections : e.g. corrections to the heat
 current and the vertex corrections, while the red curve stands for the 
same quantity but using averaged heat currents and without vertex corrections.
 The green curve  shows the scaled joint density of states. 
From the figure it is clear that as in the case of NiPd, the transition rate `${\mathbf \tau}$' is strongly dependent both on the initial and the final energies throughout the phonon frequency ($\nu$). 
 Figure \ref{com2} also clarifies that although the effect of
 disorder corrections to the current terms is small, but this effect is 
comparatively more pronounced in $Ni_{50}Pt_{50}$ than in $Ni_{50}Pd_{50}$. 
The effect of disorder corrections to current and
the vertex corrections in the present case become negligible beyond  phonon frequencies $\simeq$ 2.4 THz.
In the low frequency region, configuration fluctuation effects beyond
the mean-field is more pronounced  in NiPt than in  NiPd.
 This may be because of the two simple physical reasons : First, the NiPt is an 
alloy where both mass (m$_{Pt}$/ m$_{Ni}\simeq$ 3) as well as force constant 
(Pt-force constants are on an average 55\% larger than those of Ni) 
disorder dominates, while in NiPd alloy the mass disorder
 (m$_{Pd}$/ m$_{Ni}\simeq$ 1.812) is weaker than
NiPt and the force constants are almost the same for the two constituents.
 Second, from a purely phenomenological point of view, there is a larger 
size mismatch between Ni and Pt in NiPt alloy as compared to Ni and Pd in NiPd alloy.

\begin{figure}[t]
\vskip 0.7cm
\centering
\includegraphics[width=10cm,height=6.cm]{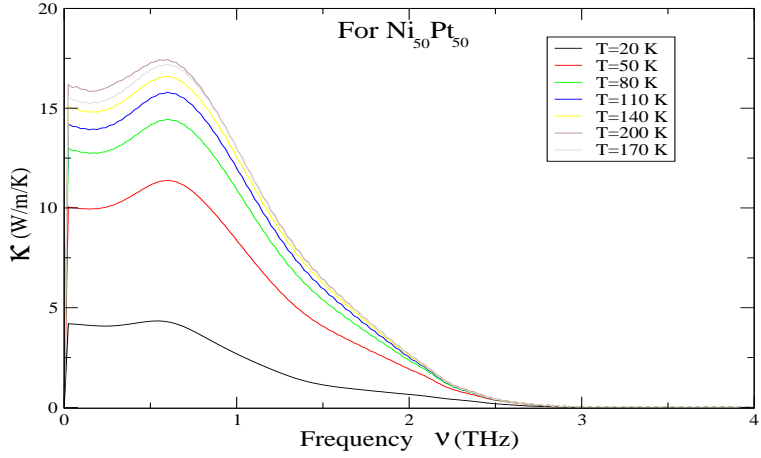}
\caption{(Color Online)\ The configuration averaged lattice thermal conductivity vs phonon frequency $\nu$ (THz) at different temperatures T for Ni$_{50}$Pt$_{50}$ alloy.}
\label{fig10}
\end{figure}

In Fig. \ref{fig10} we display the frequency dependence of lattice thermal conductivity for Ni$_{50}$Pt$_{50}$ alloy at various temperatures, evaluated using the formulae given in Eq. (\ref{eq3}) and (\ref{eq5}). We note  from Fig. \ref{fig10} that as we increase the temperature, the difference in lattice conductivity at a particular frequency is quite large in the lower temperature regime (\ $20\le T\le 80 K$), however the difference  starts to saturate as we go to higher temperatures.  As discussed in the previous section, the extrapolated value of $\kappa (\nu)$ curve from a value at $\nu >0$ (\ next to $\nu=0$\ ) to a value at $\nu=0$  is called  the d.c. value of lattice thermal conductivity ($\kappa_0$). 
Figure \ref{fig10} clearly shows that this d.c. value of the conductivity increases as we go on to increase the temperature. This is expected because in the d.c. limit the mechanism of heat conduction is transfer of energy between delocalized modes of equal energy by the heat current operator.

The behaviour of $\kappa(\nu)$ vs $\nu$ curve in both NiPd and NiPt is qualitatively similar, but there are some quantitatively different features. The maximum in NiPt is located at a lower frequency as compared with NiPd. Also the overall magntitude of lattice conductivity in NiPd is higher than NiPt.

\begin{figure}[b]
\vskip 0.6cm
\centering
\includegraphics[width=10cm,height=6.0cm]{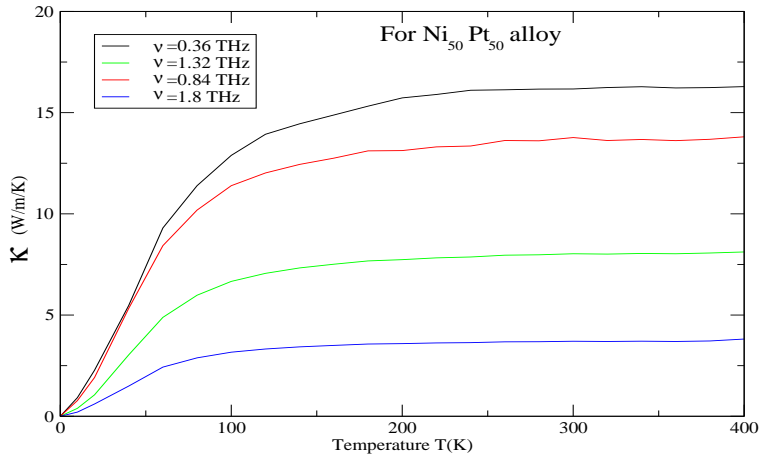}
\caption{ (Color Online)\ The averaged lattice thermal conductivity vs temperature T(K) at various cut-off frequency $\nu_{cutoff}$ for Ni$_{50}$Pt$_{50}$ alloy.}
\label{fig11}
\end{figure}
The temperature dependence of lattice conductivity for Ni$_{50}$Pt$_{50}$ alloy at various 
cut-off frequencies are shown in Fig. \ref{fig11}. The conductivity increases initially  
(\ in the low T-regime\ ) as an approximate quadratic function of temperature and 
ultimately increases smoothly to a T-independent  saturated value. As far as such 
dependence of $\kappa(T) $ in the high temperature regime is concerned, the heat in 
this conduction channel is carried by non-propagating modes which are strongly 
influenced by the disorder but mostly not localized and therefore able to conduct 
by intrinsic harmonic diffusion. This is a smooth dependence which closely resembles 
the specific heat and saturates like the specific heat at high temperaturs. Following Slack \cite{slack}
 we could call this piece the  \ ``{\it minimum thermal conductivity}". A number of authors 
have discussed that low temperature dependence of $\kappa(T)$ shows a mild plateau 
like region. In this regime the heat is mainly carried by the propagating long wavelength
 acoustic modes. The complex inelastic scattering processes then kill off the low frequency 
contribution at higher temperatures leaving a peak which becomes the plateau. 
However there are situations, where the propagating modes become reasonably well damped 
(\ as in our calculation\ ) and are no longer able to carry much heat. In such cases 
the contribution of delocalized and poorly conducting vibrations takes over, giving 
a net result in good accord with the Kittel's old idea. Under these circumstances the 
plateau like region in the low temperature regime almost disappears. The damping of 
propagating modes is also amplified as we make the alloy more and more concentrated.
 This can easily be verified by looking at the results of reference \cite{expt1}, 
which shows that as we increase the concentration the plateau like region goes down 
and gets smoother.

\begin{figure}[t]
\vskip 1cm
\centering
\includegraphics[width=10cm,height=6.cm]{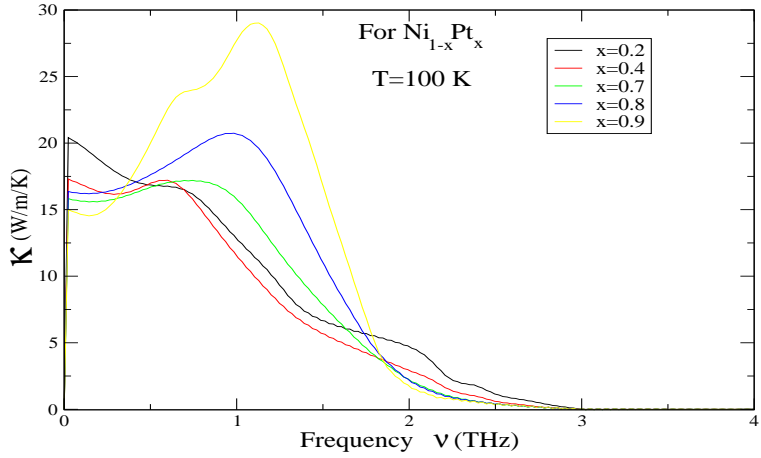}
\caption{(Color Online)\ Frequency dependence of lattice thermal conductivity for various alloys  Ni$_{1-x}$Pt$_{x}$ at T=100 K.}
\label{fig12}
\end{figure}
In Fig. \ref{fig12}, we display the frequency dependence of lattice conductivity for 
NiPt alloys at various compositions,
 but at a fixed temperature T=100 K. As before each of the curves
 have a dip at the lowest frequency. It is important to notice that as we increase the 
Pt-concentration $x$, structure appears in the behaviour of $\kappa (\nu)$.
This indicates that the structure  arises due to the contribution of Pt-atom 
(with large mass) in the alloy.

\begin{figure}[t]
\vskip 1cm
\centering
\includegraphics[width=10cm,height=6.0cm]{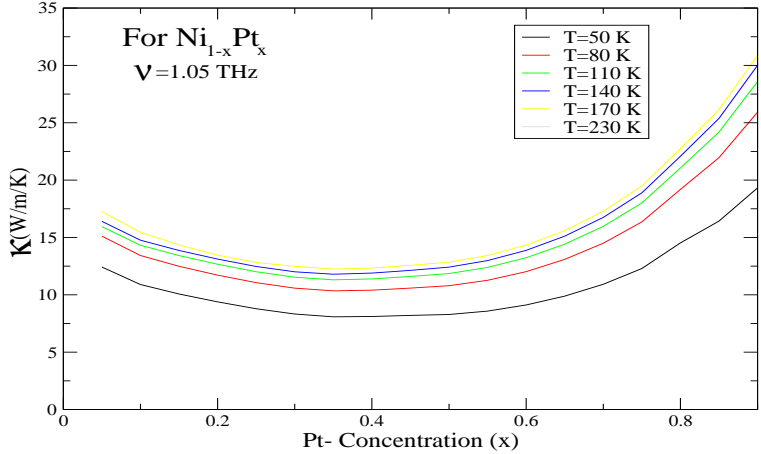}
\caption{(Color Online)\ Lattice thermal conductivity vs Pt-concentration for various temperature T at phonon frequency $\nu=1.05 THz$.}
\label{fig13}
\end{figure}

The concentration dependence of lattice conductivity at a fixed phonon frequency 
$\nu=1.05 THz$ are plotted in Fig. \ref{fig13}. The various curves in this figure stand  
for various values of the temperature T starting from a lower value of 50 K to a higher value
 of 230 K. It is clear from the figure that the concentration dependence is almost symmetric 
about x=0.5. It has been discussed by Flicker and Leath \cite{CPA} within the framework 
of coherent potential approximation that this asymmetry is a function of the size of 
the sample chosen i.e. a large N leads to less asymmetry. They have verified this statement 
by performing two calculations one for N=100 and other for N=2000. The concentration 
dependence in the later case is more symmetric as compared to the former ones. 
In our case, the results shown in Fig. \ref{fig13} are the optimal symmetric structure 
for the concentration dependence of $\kappa$. This is because our calculations are performed 
in the reciprocal space representation which involves the entire lattice.

 An interesting challenge remaining in this problem is to calculate the effect 
of adding anharmonicity to the model. The reason why one should be interested in 
calculating the anharmonicity effect is because in real systems at high temperature 
phonon-phonon Umklapp scattering becomes the dominant scattering mechanism. 
This Umklapp scattering  actually arises due to the presence of anharmonic terms
 in the Hamiltonian. The effect of this anharmonicity is to flatten 
 the lattice conductivity vs concentration curve.

\begin{figure}[t]
\centering
\includegraphics[width=13cm,height=14cm]{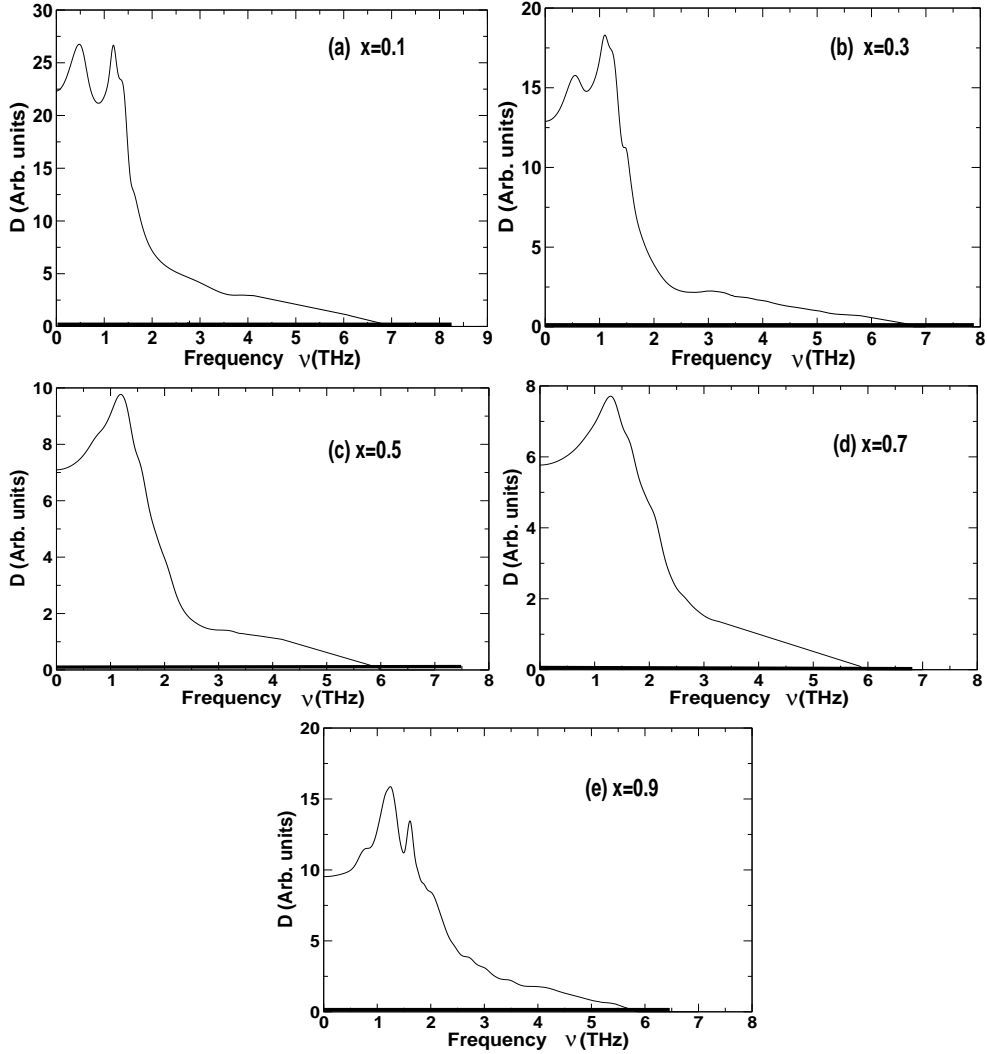}
\caption{The configuration averaged thermal diffusivities $D(\nu)$ for  Ni$_{1-x}$Pt$_{x}$ alloys. (a)\ x=0.1;\ (b)\ x=0.3;\ (c)\ x=0.5;\ (d)\ x=0.7;\ (e)\ x=0.9 .The broad line on the frequency axis shows the extent of the vibrational spectrum.}
\label{fig14}
\end{figure}

\begin{figure}
\centering
\includegraphics[width=8cm,height=9.0cm]{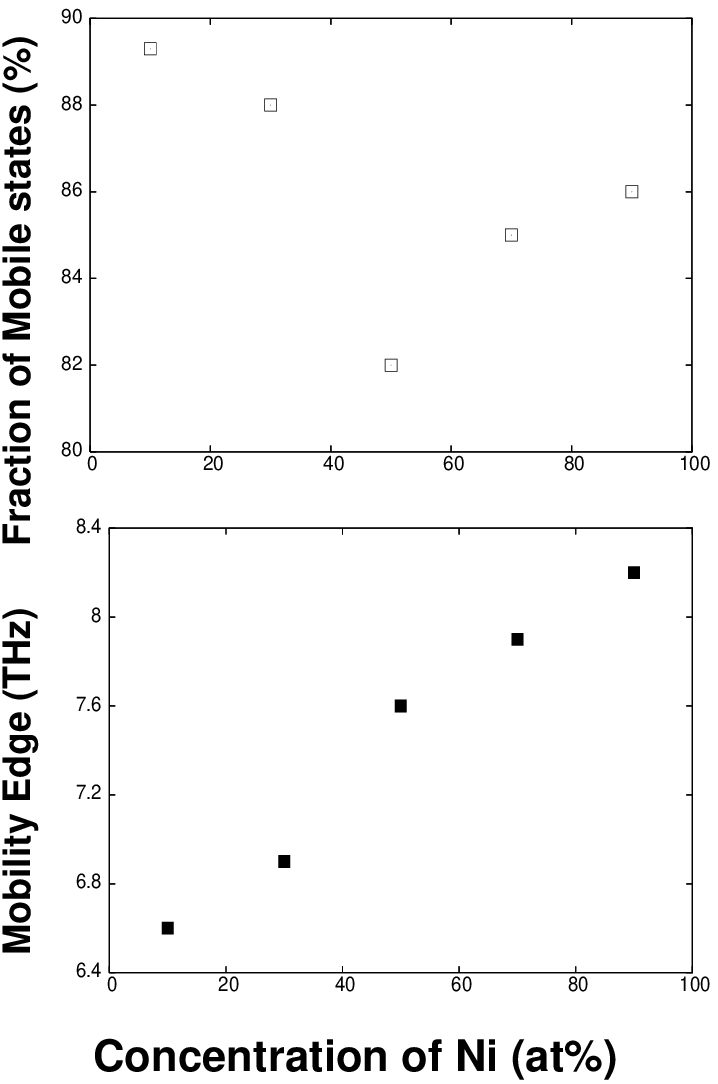}
\caption{The position of the mobility edge (bottom) and the percentage of
mobile phonon states (top) as a function of the alloy composition for NiPt alloy.}
\label{fig15}
\end{figure}

Fig. \ref{fig14} shows the frequency dependence of diffusivity $D(\nu)$ for various alloy compositions. The thick lines on the frequency axes shows the extent of the frequency spectrum. These have been obtained from the density of states calculations presented in our first paper on the subject \cite{am1}. It is clear from the  Fig. \ref{fig14} that there are basically two regions of large thermal diffusivity : one near the lower frequency region ($\simeq 0.5 THz$) and the other around a somewhat higher frequency region ($\simeq 1.25 THz$). But as we go on increasing the Pt-concentration, the former region of large diffusivity starts decreasing gradually and becomes almost flat for the maximum Pt-concentration of 90$\%$. The latter region of large diffusivity sits on a portion of the frequency spectrum  above the transverse acoustic vibrations. Here  the modes have large velocities and are probably very effective carriers of heat. The approximate linear decrease in diffusivity starts at $\simeq$ 3 THz. But the location of mobility edge in this case varies with composition in a slightly different way as compared to the case of NiPd alloy. Fig. \ref{fig15} (bottom) shows the position of the mobility edge $\nu_c$ as a function of the alloy composition. As the concentration of heavy Pt increases, the band width of the frequency spectrum (which is proportional to the square root of the mass) shrinks and the position of the mobility edge within the band also shrinks. Fig. \ref{fig15} (top) shows the fraction of the frequency band which is extended. When the disorder is the strongest, i.e. at 50-50 composition, this fraction is a minimum.

One thing is very clear from the above discussion that, in an alloy where mass disorder dominates and the forceconstant disorder is weak ( as in the case of NiPd alloy ), the complex disorder scattering processes try to localize more vibrational modes as compared to those in an alloy where both mass as well as the force constant disorder dominates ( as in the case of NiPt alloy\ ).  The result can be interpreted in a slightly different way as : \ ``{\it the role of force constant disorder in binary alloys is to make the vibrational eigenstates more delocalized }"\ , i.e. the more dominant the force constant disorder is, the more delocalized the vibrational modes will be.

\begin{figure}[b]
\vskip 0.4cm
\centering
\includegraphics[width=9.5cm,height=5.5cm]{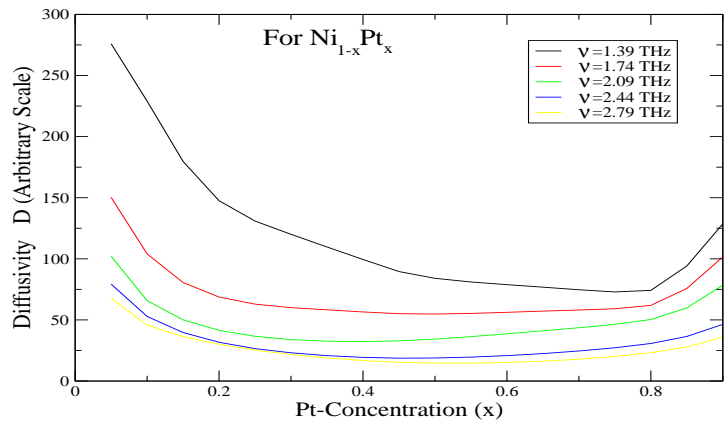}
\caption{(Color Online)\ The averaged thermal diffusivity $D(\nu)$ vs Pt-concentration at various cut-off frequencies $\nu_{cutoff}$ for Ni$_{1-x}$Pt$_{x}$ alloy. }
\label{fig16}
\end{figure}
In Fig.\ref{fig16}, we present the concentration dependence of thermal diffusivity for various phonon frequencies $\nu$. As in NiPd, the diffusivity are asymmetric about x=0.5. However this asymmetry reduces as we increase the phonon frequency. This asymmetry is reflected
also in the behaviour of thermal conductivity with alloy composition.

\section{Conclusions}
 We have performed a detailed numerical study of the theoretical formulation (AM) developed earlier by us for the lattice thermal conductivity of disordered binary alloys. We have demonstrated through our numerical results that how this multiple scattering based formalism captures the effect of off-diagonal and environmental disorder present in the problem. The use of augmented space method to keep track of the configuration of the system and the block recursion method have made the implementation simple yet powerful. This is reflected in the satisfaction of essential herglotz analytic property of the diagonal green function in our earlier calculation \cite{am1}. A significant contribution of this article beyond  the earlier theoretical approaches is the inclusion of force constant fluctuations properly in the theory. We have applied the formalism (AM) to two real disordered alloys ; namely NiPd and NiPt. We have shown that the effect of disorder corrections to the current and the vertex correction on the overall shape of lattice thermal conductivity for both the alloys are very small. Comparatively the effect is found to be more pronounced in NiPt alloy, which is due to the presence of strong disorder both in masses and force constants in this alloy. The prominence of the force constant disorder in NiPt alloy has also been demonstrated in the frequency dependence of lattice conductivity for various compositions of the Ni$_{1-x}$Pt$_{x}$ alloy. The saturation of lattice conductivity at higher temperatures has been shown for both the alloys. The concentration dependence of $\kappa$ in   Ni$_{1-x}$Pd$_{x}$ alloy has been shown to be more asymmetric about x=0.5 than in Ni$_{1-x}$Pt$_{x}$ alloy. The numerical results on the harmonic diffusivity provide an interesting idea about the localization and delocalization of the vibrational eigenstates. It says that in disordered binary alloys  \ ``\ the more stronger the force constant disorder is, the more delocalized the vibrational modes will be"\ . That is why NiPt alloy has larger fraction of delocalized states as compared to that in NiPd alloy. For both the alloys, however we had no prior information about the species dependence of the force constants but rather choose a set of force constants intuitively as we have done earlier \cite{am2,am1}. A better understanding of the role of disorder in the transport properties of random alloys could be achieved with prior information about the  force constants. These could be obtained from more microscopic theories, e.g.,\ the first principles calculation on a set of ordered alloys. Our future endeavor would be to rectify this and attempt to obtain the dynamical matrix itself from such microscopic theories.

\vskip 0.5cm
\section{\bf Acknowledgements} One of the authors (A.A.) would like to thank the Council of Scientific and Industrial Research, Govt. of India, for a fellowship during the time when this work was carried out.

\end{document}